\documentclass[12pt]{article}
\usepackage{times}
\begin{document}
\begin{titlepage}

 \hbox to \hsize{\hfil
{\bf hep-ph/9902404}}
  \vfill \large
{\bf
\begin{center}
Unitarity bound for the single-helicity-flip amplitude in elastic
$pp$-scattering.
\end{center}}
\vskip 1cm
\normalsize
\begin{center}
{\bf S. M. Troshin and N. E. Tyurin}\\[1ex] {\small  \it Institute
for High Energy Physics,\\ Protvino, Moscow Region, 142284 Russia}
\end{center}
\vskip 1.5cm
\begin{abstract}
On the basis of the $U$--matrix method of the $s$--channel
unitarization, we obtain a new unitarity bound for the 
single-helicity-flip amplitude $F_5$ of
 elastic pp-scattering at small
values of $t$.
\end{abstract}
\vfill
\end{titlepage}
The size and energy dependence of the hadron 
helicity-flip amplitude is an interesting problem in the study
of the asymptotic properties and the role of spin-dependent 
interaction and it is also
an important issue in the polarimetry
studies based on the Coulomb-Nuclear Interference (CNI) \cite{but}.
If the hadronic part of the proton-proton interaction is
helicity conserving the CNI analysing power would be then due to
the interference between a real electromagnetic helicity-flip
amplitude and an imaginary hadronic helicity-nonflip amplitude.

However, the hadronic interaction may not conserve helicity in
small angle scattering. Helicity conservation does not follow from
QCD in a region where chiral symmetry is spontaneously broken. In
Regge theory the Pomeron is usually assumed to be helicity
conserving. However this is merely an assumption which seems to be wrong;
recent experimental results on the central production of mesons
and observation of nontrivial azimuthal dependence demonstrates
the nonzero helicity tranfer by the effective Pomeron \cite{close}. 
Earlier it was shown that the unitarity generates different
 phases in the helicity-flip and nonflip Pomeron contributions 
\cite{bk} and leads to nonzero analysing power. 

The CNI
analysing power would also certainly change if there were any
nonzero  hadronic single-helicity-flip amplitude.

 The above issues are considered in the recent survey
\cite{but}. Among the new results, the most interesting one is the
unexpected bound for the single-helicity-flip amplitude $F_5$ of
  elastic $pp$-scattering, i.e.   the function
\[
\hat{F}_5(s,0)\equiv\frac{mF_5(s,t)}{\sqrt{-t}}|_{t=0}
\]
 cannot increase at
$s\rightarrow 0$ faster  than
\[cs\ln^3s,\]
while for the helicity nonflip amplitudes there is 
the Froissart-Martin bound $cs\ln^2s$.
 
In this note we show that, in fact, a stronger
bound for the function $\hat{F}_5(s,0)$ can  be obtained if
 one takes into account the unitarity in the explicit
way, e.g. uses  an unitarization method
 based on the $U$-matrix approach \cite{log}.

This method is based on the 
unitary representation for helicity amplitudes of elastic
$pp$-scattering: 
\begin{equation}
 F_{\lambda_1,\lambda_2,\lambda_3,\lambda_4}(s,b)  =
U_{\lambda_1,\lambda_2,\lambda_3,\lambda_4}(s,b)+ i\rho
(s)\sum_{\lambda ',\lambda ''} 
U_{\lambda_1,\lambda_2,\lambda ',\lambda ''}(s,b)
F_{\lambda ',\lambda '',\lambda_3,\lambda_4}(s,b),\label{heq}
\end{equation}
where $\lambda 's$ are the intial and final proton's helicities. $F_i$ are the
helicity amplitudes in the standard notations, i.e.
\[
F_1\equiv F_{1/2,1/2,1/2,1/2}, \, F_2\equiv F_{1/2,1/2,-1/2,-1/2},
\, F_3\equiv F_{1/2,-1/2,1/2,-1/2}
\]
and
\[F_4\equiv F_{1/2,-1/2,-1/2,1/2},\, F_5\equiv F_{1/2,1/2,1/2,-1/2}.
\]
The kinematical function $\rho(s)\simeq 1$ at $s\gg 4m^2$ and
will be neglected in the following. 

The functions $U_i(s,b)$
could be treated similar to the eikonal, i. e. they can
be considered as input amplitudes.

Explicit solution of Eqs. (\ref{heq}) has the following 
form:
\[
F_1(s,b)=\frac{\tilde{U}_1(s,b)[1-iU_1(s,b)]-i\tilde{U}_2(s,b)
U_2(s,b)}{[1-iU_1(s,b)]^2-[U_2(s,b)]^2},
\]
\[
F_3(s,b)=\frac{\tilde{U}_3(s,b)[1-iU_3(s,b)]-i\tilde{U}_4(s,b)
U_3(s,b)}{[1-iU_3(s,b)]^2-[U_4(s,b)]^2},
\]
where
\[
\tilde{U}_i(s,b)=U_i(s,b)+2U_5(s,b)F_5(s,b)
\]
and
\[
F_5(s,b) = \frac{U_5(s,b)}
{[1-iU_1(s,b)-iU_2(s,b)][1-iU_3(s,b)-iU_4(s,b)]-4U_5^2(s,b)}.
\]

We consider the two cases. 
First, we suppose that the helicity
nonflip functions $U_1(s,b)$ and $U_3(s,b)$ are the dominant
ones.

In this case one can get
\cite{bk}
\begin{equation}
F_{5}(s,t)=\frac{s}{\pi^2}\int_0^\infty bdb \frac{U_{5}(s,b)}
{[1-iU_{1}(s,b)][1-iU_{3}(s,b)]}J_1(b\sqrt{-t}).\label{f5}
\end{equation}

Unitarity requires that Im$U_{1,3}(s,b)\geq 0$. The functions
$U_{1,3}(s,b)$ could be different. For our purposes, however,
it is safe to assume
that they are the same $U_{1}(s,b)=U_{3}(s,b)=U(s,b)$.  For the
function $U(s,b)$ we use a simple form
\begin{equation}
U(s,b)=gs^\Delta e^{-\mu b}.\label{usb}
\end{equation}
This is a rather general parameterization for $U(s,b)$ which
provides correct analytical properties in the complex $t$--plane, i.e.
it is
 consistent with
the spectral representation  for the
function $U(s,b)$ \cite{tt}:
\begin{equation}
U(s,b)=\frac{\pi^2}{s}\int_{t_0}^\infty\rho(s,t)K_0(b\sqrt{t})dt.
\label{spect}
\end{equation}
We do not use here 
model features and do not consider detailed structure
of the helicity functions $U_i$
 but appeal to  reasonable arguments
of the general nature. To maximize the function $U_5(s,b)$ we
take it in the form $U_5(s,b)=aU(s,b)$ where $|a|<1$. 
Then  from Eq. (\ref{f5}) it follows that at
$s\rightarrow\infty$:

\begin{equation}\label{b1}
|\hat{F}_5(s,0)|\leq cs\ln^2s.
\end{equation}

This means that the magnitude of the ratio
\[
r_5(s,0)\equiv2\hat{F}_5(s,0)/[F_1(s,0)+F_3(s,0)]\] 
cannot increase with energy
and will not exceed constant at $s\rightarrow\infty$.  

This result  has a
general meaning and retains in the opposite case, i.e. in
 the case when the function $U_5(s,b)$ is a
dominant one.  We  have for the amplitude $F_5(s,t)$ the
following representation
\begin{equation}
F_{5}(s,t)=\frac{s}{\pi^2}\int_0^\infty bdb \frac{U_{5}(s,b)}
{1-4U_{5}^2(s,b)}J_1(b\sqrt{-t}).\label{f52}
\end{equation}
Using for $U_5(s,b)$ the functional dependence in the form of Eq.
(\ref{usb}) it can be easily shown that  the same  bound 
Eq. (\ref{b1})
does take place for the single-helicity-flip amplitude $\hat{F}_5$.

Thus we can state that general principles do not allow rising
behavior of $|r_5(s,0)|$. The  experimental data as well as the most
of the model predictions are consistent with
this bound \cite{trum,kr}. This result  allow us to hope
that the contribution of the single-helicity-flip amplitude
could be controllable and the effective use of the CNI
polarimeter would be possible.

 Above results were obtained in impact
parameter representation for simplicity. They can easily be
reproduced using the partial wave expansion. 

It is  accurate account of the unitarity for
the helicity amplitudes leads to Eq. (\ref{b1}), i.e.
due to unitarity the amplitude $F_5(s,b)$ has a peripheral
dependence on the variable $b$ at high energy and
\[
|F_5(s,b=0)|\rightarrow 0
\]
at $s\rightarrow\infty$. This is a consequence of the explicit
 unitarity
representation for
the helicity amplitudes and it means that the assumption on
$F_5(s,b)=constant$
at $b<R(s)$ \cite{but} appears to be inadequte (however, 
it remains to be good for the helicity-nonflip amplitudes). 

Thus,  as it was shown, we have an asymptotic bound
\[
|r_5(s,0)|\leq constant
\]
at $s\to \infty$.

To conclude, it is worth to note that only the model-dependent
estimations exist for the magnitude of the ratio $|r_5(s,0)|$,
but the rise of
the function $|r_5(s,0)|$ at $s\rightarrow \infty$ can be excluded
on the unitarity ground.

\end{document}